\newcommand{\Comment}[1]{}

\documentclass[twocolumn,a4paper]{article}
\usepackage{epsf}
\makeatletter
\def\@normalsize{\@setsize\normalsize{12pt}\xpt\@xpt
\abovedisplayskip 10pt plus2pt minus5pt\belowdisplayskip \abovedisplayskip
\abovedisplayshortskip \z@ plus3pt\belowdisplayshortskip 6pt plus3pt
minus3pt\let\@listi\@listI} 

\def\subsize{\@setsize\subsize{12pt}\xipt\@xipt}

\def\section{\@startsection {section}{1}{\z@}{11pt plus 2pt minus 2pt}
{9pt plus 2pt minus 2pt}{\bf \centering}}

\def\subsection{\@startsection {subsection}{2}{\z@}{11pt plus 2pt minus 2pt}
{9pt plus 2pt minus 2pt}{\subsize\bf}}
\makeatother

\begin{document}

\bibliographystyle{plain}

\date{}

\title{\Large\bf Online Correction of Dispersion Error in 2D Waveguide Meshes}

\author{Federico Fontana and Davide Rocchesso \\
  Dipartimento Scientifico e Tecnologico \\
  Universit\`a di Verona \\
  Strada Le Grazie 15, 37134 Verona, Italy \\
  \{fontana,rocchesso\}@sci.univr.it}
 

\maketitle

\thispagestyle{empty}

\subsection*{\centering Abstract}
{An elastic ideal 2D propagation medium, i.e., a membrane, can be simulated by models discretizing the wave equation on the time--space grid (finite--difference methods), or locally discretizing the solution of the wave equation (waveguide meshes). \Comment{Both models result in mesh structures reproducing the physical parameters of the membrane: size, shape, propagation speed of the waves, reflection properties of the edge. The two approaches provide equivalent computational structures, and introduce numerical artifacts that limit their accuracy. These artifacts are well interpreted in terms of dispersion, and cause a misalignment of the modes from their theoretical positions.} The two approaches provide equivalent computational structures, and introduce numerical dispersion that induces a misalignment of the modes from their theoretical po\-si\-tions.
Prior literature shows that dispersion can be arbitrarily reduced by oversizing and oversampling the mesh, or by adopting offline warping techniques. In this paper we propose to reduce numerical dispersion by embedding warping elements, i.e., properly tuned allpass filters, in the structure. The resulting model exhibits a significant reduction in dispersion, and requires less computational resources than a regular mesh structure having comparable accuracy.
}

\section{INTRODUCTION}
Membranes are the crucial component of most percussion instruments. Their response to an excitation, and their interaction with the rest of the musical instrument and with the environment, strongly affect the sound quality of a percussion. Physical modeling of membranes has drawn the attention of the computer music community when a new model based on the Digital Waveguide was designed, called 2-D Digital Waveguide Mesh~\cite{vanduyne932}. The model was proved to provide a computational structure equivalent to a Finite Difference \mbox{Scherme (FDS)}. In particular, it was shown that the numerical artifacts introduced by the model cause a phenomenon called dispersion. This means that, even in a flexible medium, different spatial frequency components  travel at different speeds, and this speed is direction- and frequency-dependent~\cite{Strikwerda, vanduyne932}.

Different mesh geometries have been studied: each of them have an equivalent  FDS, and exhibits its peculiar dispersion error function~\cite{FonRocIEEE2000acc}. 
The triangular geometry exhibits two valuable properties: the dispersion error is, with good approximation, independent from the direction of propagation of the spatial components; at the same time, the Triangular Waveguide Mesh (TWM) defines, from a signal-theoretic point of view, the most efficient sampling scheme among the geometries that can be derived from mesh models used in practice~\cite{Savioja2000, FonRocIEEE2000acc}.
The independency from direction has been successfully exploited~\cite{Savioja2000} to warp the signals produced by the model, using offline filtering techniques~\cite{MattiWarp}. In this paper we work on a similar idea, but warping is performed online by cascading each unit delay in the TWM with a first--order allpass filter. By properly tuning the filter parameter, we will prove that a considerable reduction of the dispersion error can be achieved in a wide range around dc.

This result is then compared with the performance of a TWM, oversized in order to reduce dispersion in the first modes. It will be shown that the warped mesh is less expensive in terms of memory and computational requirements. This evidence holds both for the straight waveguide and the FDS implementations. Our conclusion is that the most efficient, low--dispersion computational scheme for membrane modeling is a triangular FDS where the unit delays are cascaded with properly tuned allpass filters.

Having an efficient and accurate membrane model is a key step toward the construction of affordable, tunable, and realistic models of complete percussion instruments. In particular, the coupling between air and membrane~\cite{FontanaRocchessoAcustica}, and the interface with resonating structures are fundamental components that should be added to the membrane model in order to achieve better realism~\cite{AirdICMC00}.

\section{ONLINE WARPING}
For a wave traveling along the waveguide mesh, the numerical dispersion error is a function of the two spatial frequency components. In the TWM, this function is symmetric around the origin of the spatial frequency axes, with good approximation. Consequently, it makes sense to plot the dispersion factor as a single-variable function of spatial frequency, moving from the center of the surface to the absolute band edge along one of the three directions defined by the waveguide orientations\footnote{In~\cite{Savioja2000}, a function averaging the surface magnitude around the origin is constructed, resulting in a slight difference respect to the curve adopted here.}.
 \begin{figure}[t]
 \epsfxsize=220pt
 \centerline{\hfill\epsfbox{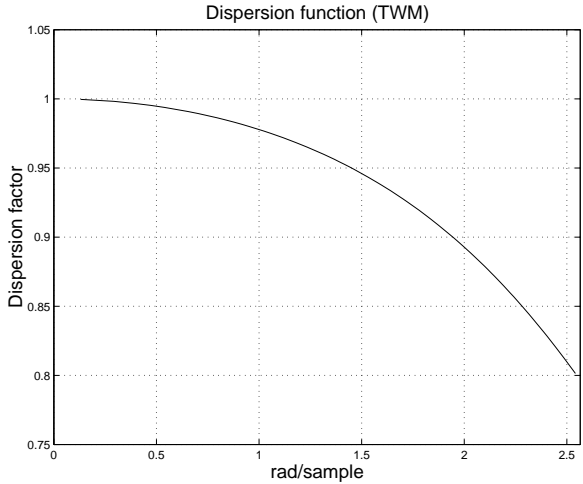}\hfill}
 \caption{Plot of the dispersion error versus temporal frequency magnitude}
 \label{figure1}
 \end{figure}
Assuming the waveguides to have unit length, the spatial band edge results to be equal to $2\pi /\sqrt{3}$ [rad$/$spatial sample]~\cite{FonRocIEEE2000acc}. A plot of the dispersion factor versus temporal frequency is then calculated recalling the nominal propagation speed factor, equal to $1/\sqrt{2}$ [spatial sample$/$time sample], affecting any finite difference model~\cite{Strikwerda}, that fixes the edge of the temporal frequency at the value $\sqrt{2}\pi /\sqrt{3}$ [rad$/$sample]. Figure~\ref{figure1} depicts a plot of the dispersion factor.

This analysis is confirmed by simulations conducted over a TWM modeling a square membrane of size $24\times 24$ waveguide sections, clamped at the four edges, excited at the central junction by an impulse. In fact, the impulse response taken at the central junction shows that the positions of its modes match well with the theoretical frequencies of the odd modes of the membrane, each one of them being shifted by its own dispersion and by the nominal propagation speed factor.
 \begin{figure}[t]
 \epsfxsize=220pt
 \centerline{\hfill\epsfbox{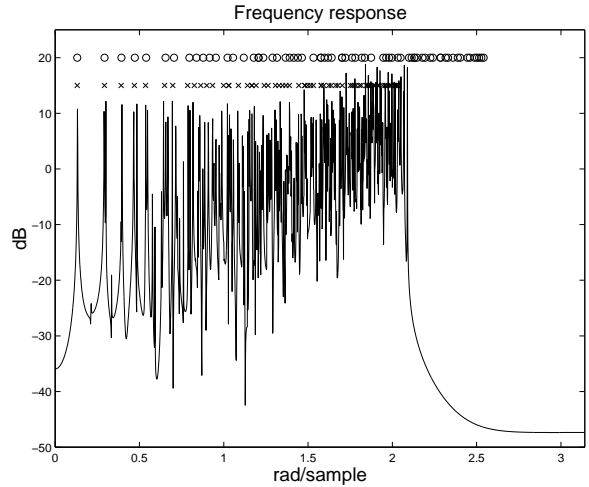}\hfill}
 \caption{Frequency response taken at the center of a TWM (size $24\times 24$) excited by an impulse at the same point. Theoretical positions of the odd modes resonating in a mebrane below $\sqrt{2}\pi /\sqrt{3}$ [rad$/$sample], weighted by the nominal propagation speed factor ($\circ$). Positions of the same modes affected by dispersion ($\times$).}
 \label{figure2}
 \end{figure}
The results are depicted in Figure~\ref{figure2}, where the frequency response of the model is plotted together with ($\circ$) the theoretical positions (compressed by the nominal propagation speed factor) of the modes resonating in the membrane below $\sqrt{2}\pi /\sqrt{3}$ [rad$/$sample], and ($\times$) the real positions of the same modes, affected by dispersion. 
Overall, dispersion introduces a modal compression that increases with frequency.
The careful reader will note a slight difference between the calculated frequency cut and the bandwidth of the signal coming from the simulation. This difference is probably due to the simplifying assumption of considering  the dispersion function as direction independent. Moreover, some modes show up as twin peaks. This may be due to the actual irregular shape of the resonator model, caused by the impossibility to design a perfectly square geometry using a TWM model. In order to conduct a controlled analytical study we avoided using interpolation along the edge, even though this is recommended in practical implementations~\cite{AirdICMC00}. 

Let $H(z)$ be the transfer function of a TWM, regardless of the excitation (input) and acquisition (output) positions. The transfer function can be handled by conformal mapping, a method consisting in the application of a particular map $T$ to the $z$-domain, in order to obtain a new, warped domain $\tilde{z}=T(z)$~\cite{Moorer83, MattiWarp}. The frequency response $H(e^{j\tilde{\omega}})$, calculated from $H(\tilde{z})|_{\tilde{z}=e^{j\tilde{\omega}}}$, changes according with the map.

Practical implementations of transfer functions obtained by conformal mapping are often affected by  non computable loops, that can sometimes be resolved~\cite{Harmaloops}. In a TWM, delay free loops appear whenever the map does not allow to ex\-tract an explicit unit delay. However, if we change  the number of unit delays  in each waveguide section of a waveguide mesh, we only change  the number of Fourier images of the frequency response\footnote{This occurs whenever a map $\tilde{z}=z^M$ is applied to a discrete-time linear filter.}, by simply compressing each single image. Now, imagine a map that translates each unit delay into the cascade of a first--order allpass filter $A(z)$ and a unit delay, $\tilde{z}^{-1}=z^{-1}A(z)$. If the allpass filter has a negative coefficient, the phase delay introduced by the filter ranges from one sample (in high frequency) to a certain value larger than one (in low frequency). Therefore, it is reasonable to expect an extra image (due to doubling the unit elements for high frequencies), and a degree of compression that decreases with increasing frequency. This is exactly the kind of behavior that is desired in order to counterbalance the effects of numerical dispersion. In order to get rid of extra frequency components it is sufficient to lowpass filter at the desired cutoff frequency, and to restore the correct positions of low-frequency partials it is sufficient to increase the temporal sampling rate.
 \begin{figure}[t]
 \epsfxsize=220pt
 \centerline{\hfill\epsfbox{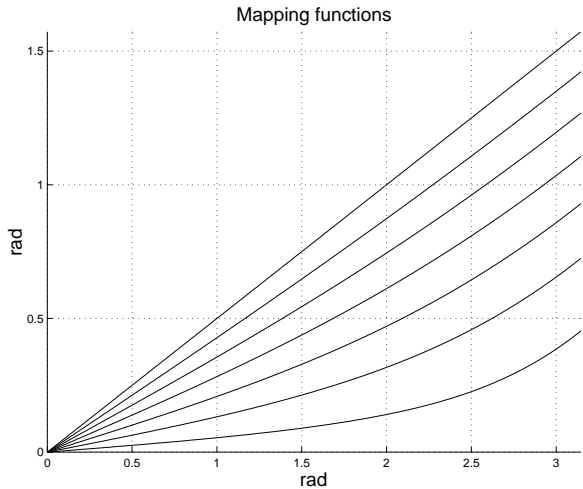}\hfill}
 \caption{Mapping functions $\tilde{z}^{-1}=z^{-1}A(z)$ for equally-spaced values of the parameter $\alpha$ of the allpass filter $A(z)$. Top line: $\alpha = 0$. Bottom line: $\alpha = -0.9$.}
 \label{figure3}
 \end{figure}

The frequency domain is warped according with the formula
\begin{equation}
\tilde{\omega}=\arctan\left[\frac{2\sin\omega\{\alpha+\cos\omega\}}{1+\alpha^2+2\alpha\cos\omega-2\sin^2\omega}\right]\:,
\end{equation}
where $\alpha$ is the parameter of the allpass filter A(z):
\begin{equation}
A(z)=\frac{\alpha+z^{-1}}{1+\alpha z^{-1}}\:.
\end{equation}
Figure~\ref{figure3} shows the mapping functions calculated for some negative values of the parameter of the allpass. The more negative is $\alpha$, the more warped are the modes, especially in the low frequency range. This result has an intuitive interpretation if we consider the phase delay of the allpass: the more negative is $\alpha$, the more delayed by the allpass filter are the lower frequencies traveling into the mesh respect to the higher ones.
 
\begin{figure}[t]
 \epsfxsize=220pt
 \centerline{\hfill\epsfbox{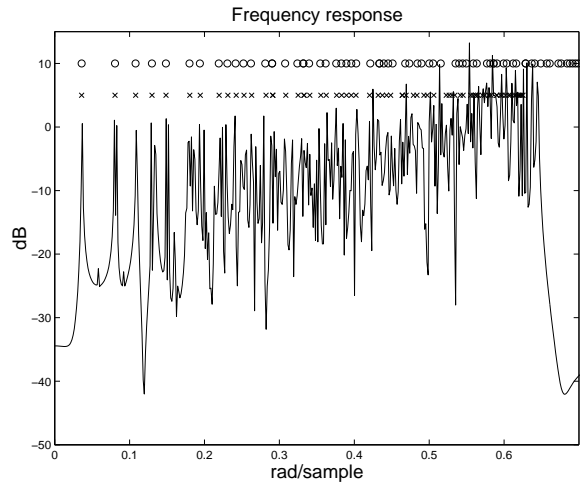}\hfill}
 \caption{Impulse response taken at the center of a warped TWM (size $24\times 24$) excited by an impulse at the same point. Theoretical positions of the odd modes resonating in a mebrane, rescaled to align the fundamentals ($\circ$). Positions of the same modes affected by residual dispersion ($\times$).}
 \label{figure4}
 \end{figure}

Choosing $\alpha=-0.45$ (curve in the middle), the warping so introduced limits the modal dispersion to very low values. Figure~\ref{figure4} shows the frequency response of the same TWM simulated before, after warping. Using the same notation of figure~\ref{figure2}, the response is compared with the ideal positions of the modes, rescaled to align the fundamentals ($\circ$), and with the real positions of the same modes, affected by the residual dispersion ($\times$). The improvement in terms of precision in the alignment of the modes is evident by comparison of crosses and circles in figures~\ref{figure2} and~\ref{figure4}.

\section{COMPUTATIONAL PERFORMANCE}
\begin{figure}[t]
 \epsfxsize=220pt
 \centerline{\hfill\epsfbox{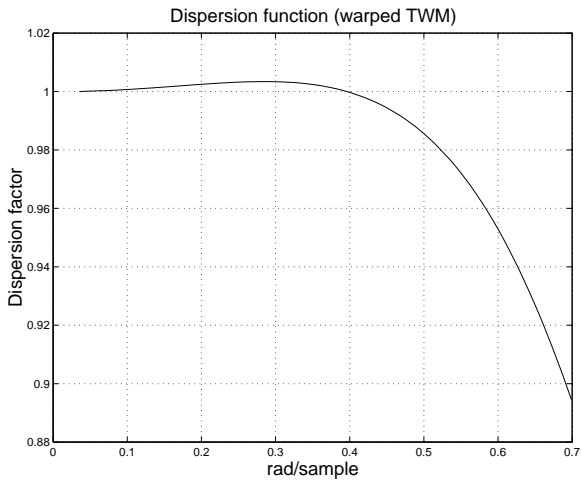}\hfill}
 \caption{Plot of the dispersion error versus temporal frequency magnitude in the warped TWM.}
 \label{figure5}
 \end{figure}

Figure~\ref{figure5} shows a plot of the dispersion factor after warping.  Dispersion is below 2\% in a range equal to 75\% of the whole band, then it climbs to the maximum.
From a perceptual viewpoint, it is not clear how much tolerance we might admit in the frequency positions of high order partials of a drum. Frequency deviation thresholds should be derived from subjective experimentation, as it was done for pia\-no sounds~\cite{scaroc}. Certainly, fig.~\ref{figure2} shows an unnatural compression of modes that results in a decrease in brightness. Moreover, the frequency distribution of resonances brings us some information about the shape of the resonating object, for instance making it possible to discriminate a circular from a square drum. The warped TWM preserves the correct distribution of modes quite well up to $75\%$ of the frequency band.

By comparison of figures~\ref{figure1} and~\ref{figure5}, we can note that a similar precision is achieved by a TWM if the waveguides are reduced to one third of the original length (the mesh is nine times denser). Then, the fundamentals can be aligned in the two models by multiplying times $1.75$ the sampling rate of the warped TWM. Finally, both the output signals must be lowpass filtered.
\begin{table}[h]
\centering
\begin{tabular}{||c||c|c|c||}                                  \hline 
                      & Sums     & Mult      & Memory          \\ \hline \hline
           TWM        &  $99$    & $9$       &  $54$           \\  \hline
           WTWM       &  $40.25$ & $22.75$   & $22.75$         \\  \hline
           FDS        &  $54$   &   $9$       &  $18$          \\  \hline
           WFDS       &  $17.5$ &   $8.75$     &  $7$          \\  \hline
\end{tabular}
\caption{Performance of the TWM (FDS) vs. warped version (`W') in terms of sums, multiplies and memory locations. Both the TWM (FDS) and its warped version allow the same dispersion tolerance.}
\label{table1}
\end{table}

From these considerations, a comparison of the TWM versus its warped version in terms of needed sums, multiplies and memory locations, based on dispersion tolerance, can be summarized in Table~\ref{table1}, where the warped models are labeled with the prefix `W', and the allpass filter is supposed to be implemented in canonical (2 multiplies, 1 delay) form. Both the straight mesh and FDS implementations are considered. The number of multiplies can be further reduced at the expense of more memory by using one-multiply allpass filter structures.

\section{CONCLUSION}
A new technique to reduce modal dispersion in a wide frequency range in TWM and triangular FDS models of 2D resonators has been presented. This technique is based on  first--order allpass filters embedded in the mesh, and it requires an increase in temporal sampling rate accompanied by lowpass filtering of the output signal. The resulting warped TWM  is shown to be less expensive in terms of computing resources and memory consumption than oversizing a TWM or FDS model. The coefficient of the embedded allpass filters is also a parameter that can be controlled to introduce tension modulation or other more exotic effects.

\bibliography{general}

\end{document}